\documentclass[11pt]{article}
\usepackage{latexsym}
\usepackage{amssymb}
\usepackage{amsmath} %%
\usepackage{epsfig}
\usepackage{ifthen}

\newtheorem{theorem}{Theorem}
\newtheorem{lemma}{Lemma}
\newenvironment{definition}{\medskip\noindent{\bf Definition:} }{\medskip}

\newenvironment{corollary}{\medskip\noindent{\bf Corollary:} }{\medskip}
\newenvironment{proof}{\noindent{\bf Proof:} }{\hfill$\Box$\medskip}

\begin{document}

\title{Deterministic Pushdown Automata and Unary Languages\footnotemark[1]~~\footnotemark[2]
\footnotetext[1]{A preliminary version of this work was presented at the 13th International Conference on Implementation and Application of Automata, CIAA 2008,
San Francisco, USA, July 21-24, 2008.%
}
\footnotetext[2]{Partially
supported by MIUR under the project PRIN ``Aspetti matematici e applicazioni emergenti degli automi e dei linguaggi formali: metodi probabilistici e combinatori in ambito di linguaggi formali''.}
}%

\author{%
Giovanni Pighizzini
\mbox{}\\
{\normalsize Dipartimento di
    Informatica e Comunicazione}\\
{\normalsize Universit\`{a} degli Studi di Milano}\\
{\normalsize via Comelico 39, 20135 Milano, Italy}\\
{\normalsize\tt pighizzini@dico.unimi.it}%
}%
\date{}%
\maketitle\thispagestyle{empty}%

\maketitle

\newcounter{lettera}
\newcounter{oldlettera}
\setcounter{oldlettera}{0}
\newenvironment{enuletter}{\begin{list}{(\alph{lettera})}%
{%\setlength{\topsep}{0cm}\setlength{\parsep}{0cm}
      \usecounter{lettera}%
      \setcounter{lettera}{\theoldlettera} 
                \setlength{\labelwidth}{20 mm}
                \setlength{\leftmargin}{10 mm}}}
{\end{list}\setcounter{oldlettera}{\thelettera}}
\newcommand{\resetletter}{\setcounter{oldlettera}{0}}

\renewcommand{\mod}{{{\rm \,mod\,}}} %%
\newcommand{\set}[1]{\{#1\}}

\newlength{\lungtop}\newlength{\lungbottom}
\newcommand{\twoargumentdash}[2]{\settowidth{\lungtop}{$\scriptstyle #1$}\settowidth{\lungbottom}{$\scriptscriptstyle #2$}\vdash\hspace{-0.7em}{\raisebox{1.1ex}{$\scriptstyle #1$}\hspace{-\lungtop}\raisebox{-0.4ex}{$\scriptscriptstyle #2$}}\ifthenelse{\lungtop>\lungbottom}{\hspace{-\lungbottom}\hspace{\lungtop}}{}}

\newcommand{\vdashstar}{\mathrel{\mbox{$\twoargumentdash{*}{}$}}}
\newcommand{\VDASH}[1]{\mathrel{\mbox{$\twoargumentdash{#1}{}$}}}
\newcommand{\vdashstarM}{\mathrel{\mbox{$\twoargumentdash{*}{M}$}}}
\newcommand{\vdashstarP}{\mathrel{\mbox{$\twoargumentdash{*}{M'}$}}}
\newcommand{\VDASHM}[1]{\mathrel{\mbox{$\twoargumentdash{#1}{M'}$}}}
\newcommand{\VDASHP}[1]{\mathrel{\mbox{$\twoargumentdash{#1}{M'}$}}}
\newcommand{\vdashM}{\mathrel{\mbox{${\VDASHM{}}$}}}
\newcommand{\vdashP}{\mathrel{\mbox{${\VDASHP{}}$}}} 

\newcommand{\genera}{\mathrel{\mbox{${\stackrel{\star}{\Rightarrow}} $}}}
\newcommand{\Genera}{\mathrel{\mbox{${\stackrel{{\scriptscriptstyle +}}{\Rightarrow}}$}}} 
\newcommand{\GENERA}[1]{\mathrel{\mbox{${\stackrel{#1}{\Rightarrow}}$}}}

\newcommand{\generaG}{\mathrel{\mbox{${\stackrel{\star}{\Rightarrow}_G}$}}}
\newcommand{\GENERAG}[1]{\mathrel{\mbox{${\stackrel{#1}{\Rightarrow}_G}$}}}

\newcommand{\generaP}{\mathrel{\mbox{${\stackrel{\star}{\Rightarrow}_{G'}}$}}} 
\newcommand{\GENERAP}[1]{\mathrel{\mbox{${\stackrel{#1}{\Rightarrow}_{G'}}$}}}

\newcommand{\exit}{{\rm exit}}
\renewcommand{\read}{{\rm read}}
\newcommand{\pop}{{\rm pop}}
\newcommand{\push}{{\rm push}}
\newcommand{\size}{{\rm size}}

\begin{abstract}
The simulation of deterministic pushdown automata defined 
over a one-letter alphabet by finite state automata is 
investigated from a descriptional complexity point of view.
We show that each unary deterministic pushdown automaton
of size $s$ can be simulated by a deterministic finite automaton
with a number of states that is exponential in $s$.
We prove that this simulation is tight. Furthermore, its
cost cannot be reduced even if it is performed
by a two-way nondeterministic automaton.
We also prove that there are unary languages for which 
deterministic pushdown automata cannot be exponentially
more succinct than finite automata. In order to state
this result, we investigate the conversion of 
deterministic pushdown automata into context-free grammars.
We prove that in the unary case the number of variables
in the resulting grammar is strictly smaller than the
number of variables needed in the case of nonunary alphabets.
\end{abstract}


\section{Introduction}
\label{sec:intro}

Deterministic context-free languages and their corresponding
devices, deterministic pushdown automata (dpda's),
have been extensively studied in the literature 
(e.g.,~\cite{GG66,Kn65,Se97,St67,Va75}). They are interesting
not only from a theoretical point of view, but even,
and perhaps mainly,
for their relevance in connection with the
implementation of efficient parsers.
It is well-known that the class of deterministic
context-free languages is a proper subclass of
that of context-free languages, 
characterized by (nondeterministic) pushdown
automata (pda's).
In the case of languages defined over a one-letter
alphabet, called \emph{unary} or \emph{tally}
languages, these classes collapse: in fact,
as proved in~\cite{GR62}, each unary context-free
language is regular. This implies that unary pda's
and unary dpda's can be simulated by finite automata.

In this paper we study the simulation of unary dpda's by
finite automata from a \emph{descriptional complexity}
point of view. As a main result, we get the cost,
in terms of the sizes of the descriptions, of the
optimal simulation between these kinds of devices.

The problem of the simulation of dpda's by finite
automata was previously studied in the literature
in the case of general 
alphabets: in~\cite{St67}
it was proved that each dpda of size $s$
accepting a regular language can be simulated by a
finite automaton with a number of states bounded
by a function which is triply exponential in $s$.
That bound was reduced to a double exponential
in~\cite{Va75}. It cannot be further reduced
because there is a matching lower bound~\cite{MF71}.

We show that in the unary case the
situation is different. In fact, we are able to prove
that each unary dpda of size $s$ can be simulated
by a one-way deterministic automaton (1dfa) with
a number of states exponential in $s$.
We prove that this simulation is tight, by showing
a family of languages exhibiting an exponential gap
between the size of dpda's accepting them, and the
number of states of equivalent 1dfa's.

As proved in~\cite{MP01}, each $n$-state
unary two-way nondeterministic finite automaton 
(2nfa) can be simulated by a 1dfa with $2^{O(\sqrt{n\log n})}$
states. This suggests the possibility of a smaller
gap between the descriptional complexities of
unary dpda's and 2nfa's. However, we show
that even in this case the gap can be exponential.

We further deepen the investigation in this subject,
in order to discover whether or not \emph{for each}
unary regular language there exists an exponential
gap between the sizes of
deterministic pushdown automata and of finite automata.
We give a negative answer to this question, by showing a
family of languages for which unary
dpda's cannot be exponentially more 
succinct than finite automata.

In order to prove this last result, we
study the problem of converting unary dpda's
into equivalent context-free grammars.
In general, given a pda with $n$ states and
$m$ input symbols, the standard conversion technique
produces an equivalent grammar with $n^2m+1$
variables. As proved in~\cite{GPW82}, this number
cannot be reduced, even if given pda is deterministic.
Here, we show that in the case of a unary alphabet,
a reduction to $2mn$ is possible.

We briefly mention that the cost of the simulation
of unary (nondeterministic) pda's by finite automata
was studied in~\cite{PSW02}, where the authors
proved that each unary pda with $n$ states and $m$
stack symbols, such that each push adds exactly one
symbol, can be simulated by a 1dfa with $2^{O(n^4m^2)}$
states. Our main result reduces this bound to
$2^{nm}$, when the given pda is deterministic.

\section{Preliminaries}
\label{sec:prel}

Given a set $S$, we let $\#S$ denote its cardinality, and $2^S$ denote the
family of all its subsets. 

A language $L$ is said to be \emph{unary} if it is defined over a 
one-letter alphabet. In this case, we let $L\subseteq a^*$. In
a similar way, an automaton is unary if its input alphabet contains
only one letter. It is easy to prove the following:

\begin{theorem}\label{th:unary}
  Let $L$ be a unary language. Then $L$ is regular if and only
  if there exist two integers $\mu\geq 0$, $\lambda\geq 1$ such that
  for each integer $n\geq\mu$, $a^n\in L$ if and only if
  $a^{n+\lambda}\in L$.
\end{theorem}
If the constant $\mu$ in Theorem~\ref{th:unary} is $0$,
then $L$ is said to be \emph{cyclic} or even
\emph{$\lambda$-cyclic}. Furthermore, in this case, 
$L$ is said to be \emph{properly $\lambda$-cyclic},
when it is not $\lambda'$-cyclic for any $\lambda'<\lambda$.
It is immediate to see that the minimum 1dfa accepting a 
properly $\lambda$-cyclic language consists of a
cycle of $\lambda$ states.

\medskip
A pushdown automaton~\cite{HU79}
$M=(Q,\Sigma,\Gamma,\delta,q_0,Z_0,F)$  is said to be 
\emph{deterministic}~\cite{GG66} if and only
if for each $q\in Q$, $Z\in\Gamma$ the following hold:
\begin{enumerate}
\item if $\delta(q,\epsilon,Z)\neq\emptyset$ 
   then $\delta(q,a,Z)=\emptyset$, for each $a\in\Sigma$, and
\item for each $\sigma\in\Sigma\cup\set{\epsilon}$, 
   $\delta(q,\sigma,Z)$ contains at most one element.
\end{enumerate}
A {\em configuration\/} of $M$ is a triple 
$(q,w,\gamma)$ where $q$ is the current state, $w$
the unread part of the input, and $\gamma$ the current content of 
the pushdown store. The leftmost symbol of $\gamma$ is the topmost stack symbol. As usual, we let $\vdash$ denote the relation between configurations
such that for two configurations $\alpha$ and $\beta$,
$\alpha\vdash\beta$ if and only if $\beta$ is reached from
$\alpha$ in one move. We also write $\alpha\VDASH{t}\beta$ if and 
only if $\beta$ can be reached from $\alpha$ in $t\geq 0$ moves, 
and $\alpha\vdashstar\beta$  if and only if $\alpha\VDASH{t}\beta$
for some $t\geq 0$.

While in the nondeterministic case acceptance by final states is
equivalent to acceptance by empty stack, for dpda's
the second condition is strictly weaker
(dpda's accepting with empty stack
characterize the class of deterministic context-free
languages having the prefix property). 
Hence, the acceptance condition we will consider in
the paper is that by \emph{final states}. In particular,
given a pda $M$, we will denote by $L(M)$ the language accepted
by it under such a condition, i.e., 
$L(M)=\set{w\in\Sigma^*\mid\exists q\in F,\gamma\in\Gamma^*: (q_0,w,Z_0)\vdashstar(q,\epsilon,\gamma)}$.

In order to simplify the exposition and the proofs of our results,
in this paper it is useful to consider pda's in a certain normal 
form~\cite{PSW02}.

\begin{enumerate}
\item At the start of the computation the pushdown store contains only the start symbol $Z_0$; this
symbol is never pushed on or popped off the stack;
\item the input is accepted if and only if the automaton reaches a final state, and all the input has been scanned;
\item if the automaton moves the input head, then no operations are performed on the stack;
\item every push adds exactly one symbol on the stack.
\end{enumerate}

The transition function $\delta$ of a pda $M$ then can be written as
\[
\delta : Q \times (\Sigma \cup \{\epsilon\}) \times \Gamma \rightarrow 2^{Q \times
(\{\read,\pop\}\cup\{\push(A) \mid A \in \Gamma\})}.
\]
In particular, for $q,p \in Q, A,B \in \Gamma, 
\sigma \in \Sigma\cup\set{\epsilon}, (p,\read) \in \delta(q,\sigma,A)$ means
that the pda $M$, in the state $q$, with $A$ at the top of the stack, by consuming the input $\sigma\in\Sigma$ or not consuming any input symbol
if $\sigma=\epsilon$,
can reach the state $p$ without changing the stack contents. $(p,\pop) \in \delta(q,\epsilon,A)$
($(p,\push(B)) \in \delta(q,\epsilon,A)$, resp.), means that
$M$, in the state $q$, with $A$ at the top of the stack, without reading any input symbol, can reach the
state $p$ by popping off the stack the symbol $A$ on the top (by pushing the symbol $B$ on the top of the stack, respectively).

It can be easily observed that each pda can be converted into an equivalent pda satisfying these conditions. Furthermore, if
the given pda is deterministic, then the resulting pda is deterministic too.
Hence, in the following we will consider dpda's in the above form.

Now, we have to introduce the measure for the size of pda's we will consider
in the paper. The literature concerning this point is very 
restricted and probably a deeper investigation should be useful.
The most extended discussion is presented in~\cite{Ha78},
where the author points out that the size of a pda $M$, denoted
as $\size(M)$,
should be defined by considering the total number of
symbols needed to write down its description and, more
precisely, the total number of symbols needed to specify its
transition function.
Converting a pda into normal form, the number of rules in the
transition function of the resulting pda is linear in the
length of the rules of the original pda, which, on the other hand,
is bounded by some constant. Hence,
the total number of symbols specifying the new pda is linear in the
total number of symbols specifying the original pda. 
Because the size of a pda in normal form is linear in the
number of rules of its transition function, and in the
deterministic case this number is linear in the product 
of the number of its states and of the number of its stack symbols, 
in the paper we will use such a product as a ``reasonable'' measure for 
the size of a dpda in normal form.

The size of a finite automaton is defined to be the number of its states.

\smallskip

A \emph{mode} of a pda $M$ is a pair belonging to $Q\times\Gamma$.
In the paper, the mode defined by a state $q$ and a symbol
$Z$ will be denoted as $[qZ]$. The \emph{mode of the configuration}
$(q,x,Z\alpha)$ is $[qZ]$.
Note that in a unary dpda, the mode of a configuration defines
the only possible move.

\smallskip

A dpda $M$ is \emph{loop-free} if and only if for each $w\in\Sigma^*$
there are $q\in Q$, $\gamma\in\Gamma^*$, $Z\in\Gamma$ such that
$(q_0,w,Z_0)\vdashstar(q,\epsilon,Z\gamma)$ and 
$\delta(q,\epsilon,Z)=\emptyset$, i.e., for each input
string the computation cannot enter in an infinite loop of
$\epsilon$-moves. It is known that each dpda can be converted into
an equivalent loop-free dpda~\cite{GG66}. 
In the unary case such a conversion can be done without
increasing the size of the given dpda. 
In fact, we can write a procedure that given a mode $[qA]$ 
simulates the $\epsilon$-moves of $M$ in order to make a list of 
the modes  reachable
from the configuration $(q,\epsilon,A)$. If a mode is visited twice,
then the computation enters a loop. In this case, the transition 
function of $M$ can be modified
by setting $\delta(p,\epsilon,B)=\emptyset$ 
for each mode visited in the
simulation. Note that the procedure ends before $\size(M)$ steps.
Hence, in the following, without loss of generality, we will 
suppose that each unary dpda we consider is loop-free.

\section{Simulation of unary dpda's by finite automata}
\label{sec:simulation}

In this section we prove our main result: in fact we show
that each unary dpda $M$ can be simulated by a 1dfa
whose number of states is exponential in the size 
of $M$. We will also show that this simulation is tight.

Let us consider a given unary dpda $M$. We start by
introducing some useful notions and lemmas:

\begin{definition}
Given two modes $[qA]$ and $[pB]$, we define
$[qA]\leq [pB]$ if and only if there are integers
$k,h\geq 0$ and strings $\alpha,\beta\in\Gamma^*$, such that:
\begin{itemize}
\item $(q_0,a^k,Z_0)\vdashstar(q,\epsilon,A\alpha)$,
      $(q,a^h,A)\vdashstar(p,\epsilon,B\beta)$, and
\item if $(q_0,a^{k'},Z_0)\vdashstar(p,\epsilon,B\beta')$
      for some $k'<k$, $\beta'\in\Gamma^*$, then there is an integer $k''$
      with $k'+k''<k$ and a state $p'\in Q$, such that
      $(p,a^{k''},B)\vdashstar(p',\epsilon,\epsilon)$.
\end{itemize}
\end{definition}
Intuitively, $[qA]\leq[pB]$ means that $M$ from the initial 
configuration can reach a configuration with mode $[qA]$
by a computation $(q_0,a^k,Z_0)\vdashstar(q,\epsilon,A\alpha)$ and,
after that, it can reach a configuration with mode $[pB]$ by
a computation which does not use the portion of the
stack below $A$, i.e., the portion containing $\alpha$.
Furthermore, if during the computation 
$(q_0,a^k,Z_0)\vdashstar(q,\epsilon,A\alpha)$ a configuration
with mode $[pB]$ and stack height $h$ is reached, 
then in some subsequent step of the same computation
the stack height must decrease below height $h$. In other words,
for all integers $k'$ and $k''$ with $k'+k''=k$, it is not
possible that 
$(q_0,a^{k'},Z_0)\vdashstar(p,\epsilon,B\beta')$ and
$(p,a^{k''},B)\vdashstar(q,\epsilon,A\alpha')$, for
some $\alpha',\beta'\in\Gamma^*$.

\begin{lemma}\label{lemma:order}
The relation $\leq$ defines a partial
order on the set of the modes.
\end{lemma}
\begin{proof}
Clearly, the relation $\leq$ is reflexive.
To prove that it is antisymmetric, we consider two modes
$[qA]$ and $[pB]$ and we show that $[qA]\leq [pB]$
and $[pB]\leq [qA]$ imply $[qA]=[pB]$.

By definition of $\leq$, for suitable integers $k,h,s,t$, 
and strings $\alpha,\beta,\eta,\gamma\in\Gamma^*$, we have:
\resetletter
\begin{enuletter}
\item $(q_0,a^k,Z_0)\vdashstar(q,\epsilon,A\alpha)$,
\item $(q,a^h,A)\vdashstar(p,\epsilon,B\beta)$,
\item $(q_0,a^s,Z_0)\vdashstar(p,\epsilon,B\gamma)$,
\item $(p,a^t,B)\vdashstar(q,\epsilon,A\eta)$.
\end{enuletter}

Considering (b) and (d), we can observe that 
when $M$ reaches a configuration with the mode $[qA]$ 
($[pB]$, respectively), the symbol
$A$ ($B$, resp.) will never be popped off the stack,
i.e.:
\begin{enuletter}
\item for each $n\geq 0$, there are $q',p'\in Q$,
      $\alpha',\beta'\in\Gamma^*$ such that: 
      $(q,a^n,A)\vdashstar(q',\epsilon,\alpha'A)$
      and
      $(p,a^n,B)\vdashstar(p',\epsilon,\beta'B)$.      
\end{enuletter}
We now suppose that $s\neq k$. If $s<k$ then
from (c) and (a) we get:
\begin{enuletter}
\item $(q_0,a^k,Z_0)\vdashstar(p,a^{k-s},B\gamma)
      \vdashstar(q,\epsilon,A\alpha)$.
\end{enuletter}
By the definition of $\leq$, this implies the
existence of an integer $l$ with $s+l<k$ and a
state $p''$ such  that
$(p,a^l,B)\vdashstar(p'',\epsilon,\epsilon)$,
which is a contradiction to (e). In a symmetrical
way, by supposing $k<s$, we get a contradiction.

This permits us to conclude that $s=k$ and hence that
$[qA]=[pB]$.

\bigskip

We now prove that $\leq$ is transitive. To this aim
we suppose that
$[qA]\leq[pB]$ and $[pB]\leq[rC]$ and we show that
$[qA]\leq[rC]$.
If $[pB]=[rC]$ then the result is trivial.
Hence, from now on, we suppose $[pB]\neq[rC]$.

We consider integers $k,h,s,t\geq 0$ and
strings $\alpha,\beta,\eta,\gamma\in\Gamma^*$ such that:
\resetletter
\begin{enuletter}
\item $(q_0,a^k,Z_0)\vdashstar(q,\epsilon,A\alpha)$,
\item $(q,a^h,A)\vdashstar(p,\epsilon,B\beta)$,
\item $(q_0,a^s,Z_0)\vdashstar(p,\epsilon,B\eta)$,
\item $(p,a^t,B)\vdashstar(r,\epsilon,C\gamma)$.
\end{enuletter}
From (b) and (d) we get:
\begin{enuletter}
\item $(q,a^{h+t},A)\vdashstar(r,\epsilon, C\gamma\beta)$.
\end{enuletter}
Suppose, by contradiction, that $[qA]\leq[rC]$ does not hold.
Considering the definition of $\leq$, (a) and (e), it turns
out that it must exist two integers $k_1$ and $k_2$ with
$k_1+k_2=k$ such that:
\begin{enuletter}
\item $(q_0,a^{k_1},Z_0)\vdashstar(r,\epsilon,C\gamma_1)$ and
\item $(r,a^{k_2},C)\vdashstar(q,\epsilon,A\gamma_2)$
\end{enuletter}
with $\alpha=\gamma_2\gamma_1$. 
From (g) and (b) we get:
\begin{enuletter}
\item $(r,a^{k_2+h},C)\vdashstar(p,\epsilon,B\beta\gamma_2)$
\end{enuletter}
Because $[rC]\neq[pB]$ and $[pB]\leq[rC]$, it turns out that
$[rC]\leq[pB]$ cannot hold. Considering (f) and (h) this
implies the existence of two integers $k'$ and $k''$
with $k'+k''=k_1$ such that
\begin{enuletter}
\item $(q_0,a^{k'},Z_0)\vdashstar(p,\epsilon,B\gamma')$
\item $(p,a^{k''},B)\vdashstar(r,\epsilon,C\gamma'')$
\end{enuletter}
with $\gamma''\gamma'=\gamma_1$. Hence:
\begin{enuletter}
\item $(p,a^{k''+k_2},B)\vdashstar(q,\epsilon,A\gamma_2\gamma'')$
\end{enuletter}
But this, together with (i), gives a contradiction to the hypothesis that
$[qA]\leq[pB]$. Hence, we are finally able to conclude
that $[qA]\leq[rC]$.

\end{proof}

A configuration completely describes the status of a pda in a given
instant and gives enough information to simulate the remaining steps
of a computation. However, in order to study the properties of
the computations of dpda's, it is useful to have a richer description,
which also takes into account the states reached in some previous
computation steps. To this aim we now introduce the notion of
history.  Before doing that, we observe that the next move from
a configuration of a unary dpda depends only on the current
mode. If such a move requires the reading of an input symbol and
all the input has been consumed, then the computation stops.
Hence, given a unary dpda $M$, for each integer $t$ there 
exists at most one configuration that can be reached after 
$t$ computation steps. Such a configuration
will be reached  if the input is long enough.

\begin{definition}
  For each integer $t\geq 0$, the \emph{history} $h_t$ of $M$
  at the time $t$ is a sequence of modes
  $[q_mZ_m][q_{m-1}Z_{m-1}]\cdots[q_1Z_1]$ such that:
  \begin{itemize}
  \item $Z_mZ_{m-1}\cdots Z_1$ is the content of the stack
    after the execution of $t$ transitions from the initial configuration,
  \item for each integer $i$, $1\leq i\leq m$, $[q_iZ_i]$ was the mode
    of the last configuration having stack height $i$, in the computation
    $(q_0,x,Z_0)\VDASH{t}(q_m,\epsilon,Z_mZ_{m-1}\cdots Z_1)$,
    for a suitable $x\in a^*$.
  \end{itemize}
  The \emph{mode at the time} $t$, denoted as $m_t$, is
  the leftmost symbol of $h_t$, i.e., the pair representing
  the state and the stack top of $M$ after $t$ transitions.\footnote{
  Because the start symbol $Z_0$ is never popped off the stack,
  actually we can observe that in each history the
  symbol $Z_1$ of the rightmost mode coincides with $Z_0$.}
\end{definition}
In what follows we let $H$ denote the set of all histories of
$M$, i.e., $H=\set{h_t\mid t\geq 0}$.

\begin{lemma}\label{lemma:history}
  Let $h_t=[q_mZ_m][q_{m-1}Z_{m-1}]\cdots[q_1Z_1]$ be the
  history at the time $t$, for a given $t\geq 0$. Then:
  \begin{enumerate}
  \item\label{history:1}
  For $i=1,\ldots,m-1$, there is an integer $t_i$ s.t.\
  $h_{t_i}=[q_iZ_i][q_{i-1}Z_{i-1}]\cdots[q_1Z_1]$,
  $(q_i,x,Z_i)\vdashstar(q_{i+1},\epsilon,Z_{i+1}Z_i)$, 
  for some $x\in a^*$, and $h_{t_i}$ is a suffix of
  each $h_j$, for each integer $j$ such that $t_i<j\leq m$.
  Furthermore $0\leq t_1<t_2<\cdots<t_{m-1}<t$.
  
  \item\label{history:2}
  If all the modes in $h_t$ are different then
  $[q_1Z_1]\leq\cdots\leq[q_mZ_m]$.

  \item\label{history:3}
  If $h_{\mu}=h_{\mu+\lambda}$ for some $\mu\geq 0$, $\lambda\geq 1$, then
  $h_{\mu+i}=h_{\mu+\lambda+i}$, for each $i\geq 0$.

  \end{enumerate}
\end{lemma}
\begin{proof}
For each $i$, $1\leq i\leq m$, let $t_i\geq 0$ be the
largest integer such that $|h_{t_i}|=i$. (Note that $t_m=t$.)

Hence, the stack height at each step $j$, $t_i<j\leq m$, must
be greater than $i$. This implies that the first $i$ symbols on
the stack cannot be modified after step $t_i$, i.e., 
$h_{t_i}=[q_iZ_i]\cdots[q_1Z_1]$,
and, in the case $i<m$, 
$(q_i,x,Z_i)\vdashstar(q_{i+1},\epsilon,Z_{i+1}Z_i)$, for some
input $x$. Hence, (\ref{history:1}) easily follows.

\medskip
To prove (\ref{history:2}), we also observe that
$(q_0,a^k,Z_0)\vdashstar(q_i,\epsilon,Z_iZ_{i-1}\cdots Z_1)$, 
for some $k\geq 0$.
Suppose that $[q_iZ_i]\leq[q_{i+1}Z_{i+1}]$ is not true. Hence,
$(q_0,a^{k'},Z_0)\vdashstar(q_{i+1},\epsilon,Z_{i+1}\gamma')$ and
$(q_{i+1},a^{k''},Z_{i+1})\vdashstar(q_i,\epsilon,Z_i\gamma''Z_{i+1})$ 
for some $k',k''$, with $k'+k''=k$ and $\gamma',\gamma''\in\Gamma^*$.
Thus, $Z_i\gamma'' Z_{i+1}\gamma'=Z_i\cdots Z_1$ and 
$[q_{i+1}Z_{i+1}]=[q_jZ_j]$ for some $j<i$, which is a contradiction
to the initial hypothesis that $h_t$ does not contain any repetition.

Hence, we get that $[q_iZ_i]\leq[q_{i+1}Z_{i+1}]$ and
(\ref{history:2}) follows by Lemma~\ref{lemma:order}.

\medskip
To prove (\ref{history:3}), we observe that $h_{\mu}=h_{\mu+\lambda}$
implies that the configurations reached at the steps $\mu$ and
$\mu+\lambda$ coincide. Since $M$ is unary and deterministic, it
immediately follows that for each $i>0$ at
steps $\mu+i$ and $\mu+\lambda+i$ the same move is performed.
Hence, $h_{\mu+i}=h_{\mu+\lambda+i}$.
\end{proof}

\begin{lemma}\label{lemma:H}
  The set $H$ contains infinitely many histories if and only if
  there exist two integers $\mu\geq 0$, $\lambda\geq 1$, and
  $\lambda$ nonempty sequences of modes 
  $\tilde{h}_1,\ldots,\tilde{h}_{\lambda}$,
  such that
  \[
  h_{\mu+1}=\tilde{h}_1h_{\mu},~ 
  h_{\mu+2}=\tilde{h}_2h_{\mu},~\ldots,~ 
  h_{\mu+k\lambda+i}=\tilde{h}_i(\tilde{h}_{\lambda})^kh_{\mu},
  \]
  for all integers $k\geq 0$, $0\leq i<\lambda$.
  
  Furthermore, if such $\mu$ and $\lambda$ exist then their sum
  does not exceed $2^{\#Q\cdot\#\Gamma}$, while if $H$ is finite
  then its cardinality is less than $2^{\#Q\cdot\#\Gamma}$.
\end{lemma}
\begin{proof}
  Suppose that $H$ contains infinitely many elements, and
  consider the smallest index $t$ such that the history
  $h_t=[q_mZ_m]\cdots[q_1Z_1]$ contains a repetition.
  In the light of Lemma~\ref{lemma:history}(\ref{history:1}),
  the mode $[q_mZ_m]$ must be repeated in $h_t$, namely
  there is an index $i$, $0\leq i<m$, such that $[q_mZ_m]=[q_iZ_i]$,
  an integer $\mu$, $1\leq\mu<t$, such that 
  $h_{\mu}=[q_iZ_i]\cdots[q_1Z_1]$,
  and some sequences $\tilde{h}_1,\ldots,\tilde{h}_{\lambda}$,
  where $\lambda=t-\mu$,
  such that
  $h_{\mu+1}=\tilde{h}_1h_{\mu}$, \ldots,
  $h_{\mu+\lambda}=\tilde{h}_{\lambda}h_{\mu}$.
  Note that the sequences $\tilde{h}_{i}$ cannot be empty
  (otherwise, by Lemma~\ref{lemma:history}(\ref{history:3}),
  $H$ cannot contain infinitely many elements).
  Because the transitions after time $\mu$  depend only on the
  mode $[q_iZ_i]$ and on the modes in the sequences
  $\tilde{h}_1,\ldots,\tilde{h}_{\lambda}$, and the
  mode at the time $\mu+\lambda=t$ is $[q_iZ_i]$, then it is
  not difficult to conclude that 
  $h_{\mu+\lambda+1}=\tilde{h}_1\tilde{h}_{\lambda}h_{\mu}$,
  $h_{\mu+\lambda+2}=\tilde{h}_2\tilde{h}_{\lambda}h_{\mu}$,
  \ldots
  $h_{\mu+k\lambda+i}=\tilde{h}_i(\tilde{h}_{\lambda})^kh_{\mu}$,
  for $k\geq 0$, $0\leq i<\lambda$.

  The converse is trivial.
  
  Finally, we observe that, by Lemma~\ref{lemma:history}(\ref{history:2}),
  the sets of modes belonging to two different histories $h_t$ and $h_{t'}$ 
  not containing any repetition must be different.
  This implies that the
  number of histories without repetitions does not exceed the
  number of all possible nonempty sets of modes, i.e., it is
  at most $2^{\#Q\cdot\#\Gamma}-1$. Hence, if the
  history $h_{2^{\#Q\cdot\#\Gamma}}$ does not contain any
  repetition, then it coincides with some history $h_t$,
  for a $t<2^{\#Q\cdot\#\Gamma}$. By 
  Lemma~\ref{lemma:history}(\ref{history:3}) this implies
  that $H$ is finite.
\end{proof}

\begin{lemma}\label{lemma:modes}
  The sequence $(m_t)_{t\geq 0}$ is ultimately periodic.
  More precisely, there are integers $\mu\geq 0,\lambda\geq 1$
  such that $\mu+\lambda\leq 2^{\#Q\#\Gamma}$ and
  $m_t=m_{t+\lambda}$, for each $t\geq\mu$.
\end{lemma}
\begin{proof}
  By Lemma~\ref{lemma:history}(\ref{history:3}), if $H$ is finite
  then $(h_t)_{t\geq 0}$ is ultimately periodic, and hence
  even $(m_t)_{t\geq 0}$ is ultimately periodic. Note that,
  as a consequence of Lemma~\ref{lemma:H}, in this case
  the set $H$ cannot contain more than $2^{\#Q\#\Gamma}-1$ elements.
  This gives the upper bounds on $\mu+\lambda$.
  
  If $H$ is infinite then the sequence of histories $(h_t)_{t\geq 0}$ 
  is not periodic. However, the sequence of modes $(m_t)_{t\geq 0}$
  is defined by the leftmost symbols of $(h_t)_{t\geq 0}$.
  Hence, by Lemma~\ref{lemma:H}, it is periodic, with 
  $\mu+\lambda\leq 2^{\#Q\#\Gamma}$.
\end{proof}

Now, we are ready to prove our main result:

\begin{theorem}\label{th:simulation}
  Let $L\subseteq a ^*$ be accepted by a dpda $M$ in normal
  form with $n$ states and $m$ stack symbols. Then $L$
  is accepted by a 1dfa with at most $2^{mn}$ states.
\end{theorem}
\begin{proof}
  The acceptance or rejection of a word depends only on the states
  that are reached by consuming it (and possibly performing some
  $\epsilon$-moves). By Lemma~\ref{lemma:modes} the sequence of 
  the modes that can be reached in computation steps is ultimately
  periodic. This implies that also the sequence of the reached
  states, which gives the acceptance or the rejection, 
  is ultimately periodic.
  Hence, it is possible to build a 1dfa accepting the language.
  The upper bound on the number of the states derives from
  Lemma~\ref{lemma:modes}. 
\end{proof}

As a consequence of Theorem~\ref{th:simulation}, each unary dpda $M$
of size $s$ can be simulated by a 1dfa with a number of states
exponential in $s$. We now prove that such a simulation is optimal.
In particular, we show that for each integer $s$ there exists
a language which is accepted by a dpda of size $O(s)$ such
that any equivalent 1dfa needs $2^s$ states.

More precisely, for each integer $s$, we consider the set
of the multiples of $2^s$, written in unary notation, namely
the language $L_s=\set{a^{2^s}}^*$.

Given $s>0$, we can build a dpda accepting $L_s$ that, 
from the initial configuration,
reaches a configuration with the state $q_0$ and the pushdown
containing only $Z_0$,
every time it consumes an input factor of length $2^s$, i.e.,
$(q_0,a^{2^s},Z_0)\vdashstar(q_0,\epsilon,Z_0)$.
The state $q_0$ is the only final state and it cannot be reached
in the other steps of the computation.
The computation from $(q_0,a^{2^s},Z_0)$ to 
$(q_0,\epsilon,Z_0)$ uses a procedure that, given an integer
$i$, consumes $2^i$ input symbols. For $i>0$ the procedure makes
two recursive calls, each one of them consuming $2^{i-1}$
symbols. In the implementation, two stack symbols $A_{i-1}$ and $B_{i-1}$
are used, respectively, to keep track of the first and of the second
recursive call of the procedure.
For example, for $s=3$, a configuration with the pushdown store
containing $B_0A_1B_2Z_0$ will be reached after consuming
$2^2+2^0$ input symbols and performing some $\epsilon$-moves. 
The formal definition is below:
\begin{itemize}
\item $Q=\set{q_0,q_1,q_2,q_3}$
\item $\Gamma=\set{Z_0,A_0,A_1,\ldots,A_{s-1},B_0,B_1,\ldots,B_{s-1}}$
\item $\delta(q_0,\epsilon,Z_0)=\set{(q_1,\push(A_{s-1}))}$\\
      $\delta(q_1,a,A_0)=\set{(q_3,\read)}$\\
      $\delta(q_1,a,B_0)=\set{(q_3,\read)}$\\
      $\delta(q_1,\epsilon,A_i)=\delta(q_1,\epsilon,B_i)=
      \set{(q_1,\push(A_{i-1}))}$, for $i=1,\ldots,s-1$\\
      $\delta(q_2,\epsilon,A_i)=\delta(q_2,\epsilon,B_i)=
      \set{(q_1,\push(B_{i-1}))}$, for $i=1,\ldots,s-1$\\      
      $\delta(q_3,\epsilon,A_i)=\set{(q_2,\pop)}$, for
      $i=0,\ldots,s-1$\\ 
      $\delta(q_3,\epsilon,B_i)=\set{(q_3,\pop)}$, for
      $i=0,\ldots,s-1$\\ 
      $\delta(q_2,\epsilon,Z_0)=\set{(q_1,\push(B_{s-1}))}$\\
      $\delta(q_3,\epsilon,Z_0)=\set{(q_0,Z_0)}$
\item $F=\set{q_0}$.
\end{itemize}

\begin{theorem}\label{th:tight}
  For each integer $s>0$, the language $L_s$ is accepted by a
  dpda of size $8s+4$ but the minumum 1dfa accepting it
  contains exactly $2^s$ states.
\end{theorem}
\begin{proof}
  First, we prove by induction on $i=0,\ldots,s-1$, that
  $(q_1,a^{2^i},A_i)\vdashstar(q_2,\epsilon,\epsilon)$
  and $(q_1,a^{2^i},B_i)\vdashstar(q_3,\epsilon,\epsilon)$.
  The basis, $i=0$, is trivial.
  For $i>0$ the computations, obtained using the induction
  hypothesis, are the following, where the symbol $C$ can
  be replaced by $A_i$ and by $B_i$:
  \begin{eqnarray*}
  (q_1,a^{2^i},C)\vdash(q_1,a^{2^i},A_{i-1}C)\vdashstar
  (q_2,a^{2^{i-1}},C)\vdash(q_1,,a^{2^{i-1}},B_{i-1}C)
  \vdashstar(q_3,\epsilon,C).
  \end{eqnarray*}
  and the last step is 
  $(q_3,\epsilon,A_i)\vdash(q_2,\epsilon,\epsilon)$
  or 
  $(q_3,\epsilon,B_i)\vdash(q_3,\epsilon,\epsilon)$.  
  
  As a consequence, the dpda of size $8s+4$ defined above
  recognizes $L_s$.
  Because $L_s$ is properly $2^s$-cyclic,  the minimum 
  1dfa accepting it has $2^s$ states.
\end{proof}

Using Theorem~9 of~\cite{MP00}, it is possible to
prove that also any 2nfa accepting the language $L_s$ 
must have at least $2^s$ states. Hence we get the following:

\begin{corollary}
  Unary determistic pushdown automata can be exponentially
  more succinct than two-way nondeterministic finite automata.
\end{corollary}

\section{Unary dpda's and context-free grammars}
\label{sec:grammar}

In this section we study the conversion of unary dpda's
into context-free grammars. Given a pda with $n$ states
and $m$ stack symbols, the standard conversion produces
a context-free grammar with $n^2m+1$ variables.
In~\cite{GPW82} it has been proved that such a number
cannot be reduced, even if the given pda is deterministic.
As we prove in this section, in the unary case the
situation is different. In fact, we show how to get a 
grammar with $2nm$ variables.
This transformation will be useful in the last part of 
the paper to prove the existence of languages
for which dpda's cannot be exponentially more succinct than
1dfa's.

\medskip

Let $M=(Q,\set{a},\Gamma,\delta,q_0,Z_0,F)$ be a unary
dpda in normal form.

First of all, we observe that for each mode $[qA]$ there 
exists at most one state $p$ such that 
$(q,x,A)\vdashstar(p,\epsilon,\epsilon)$ for some
$x\in a^*$. We denote such a state by $\exit[qA]$ and 
we call the sequence of moves from $(q,x,A)$ to 
$(p,\epsilon,\epsilon)$, the \emph{segment of computation}
from $[qA]$. Note that given two modes
$[qA]$ and $[q'A]$, if $(q,x,A)\vdashstar(q',\epsilon,A)$,
for some $x\in a^*$, then $\exit[qA]=\exit[q'A]$.

\medskip
We now define a grammar $G=(V,\set{a},P,S)$ and
we will show that it is
equivalent to $M$. The set of variables is
$V=Q\times\Gamma\times\set{0,1}$. The
elements of $V$ will be denoted as
$[qA]_b$, where $[qA]$ is a mode and
$b\in\set{0,1}$. The start symbol of
the grammar is $S=[q_0Z_0]_1$.

The productions of $G$ are defined in order to derive
from each variable $[qA]_0$ the string $x$ consumed in 
the segment of computation from $[qA]$, and from
each variable $[qA]_1$ 
all the strings $x$ such that $M$, from
a configuration with mode $[qA]$ can reach
a final configuration, consuming $x$, before
completing the segment from $[qA]$. They are listed
below, by considering the possible moves of $M$:
\begin{itemize}
\item \emph{Push moves:}
  For $\delta(q,\epsilon,A)=\set{(p,\push(B))}$,
  there is the production
  \resetletter
  \begin{enuletter}
    \item $[qA]_1\rightarrow[pB]_1$
  \end{enuletter}
  Furthermore, if $\exit[pB]$ is defined, with
  $\exit[pB]=q'$, then there are the productions
  \begin{enuletter}
    \item $[qA]_0\rightarrow[pB]_0[q'A]_0$
    \item $[qA]_1\rightarrow[pB]_0[q'A]_1$
  \end{enuletter}
\item \emph{Pop moves:}
  For $\delta(q,\epsilon,A)=\set{(p,\pop)}$,
  there is the production
  \begin{enuletter}
    \item $[qA]_0\rightarrow\epsilon$
  \end{enuletter}    
\item \emph{Read moves:}
  For $\delta(q,\sigma,A)=\set{(p,\read)}$,
  with $\sigma\in\set{\epsilon,a}$, and for
  each $b\in\set{0,1}$, there is the production
  \begin{enuletter}
    \item $[qA]_b\rightarrow\sigma[pA]_b$
  \end{enuletter}    
\item \emph{Acceptance:}
  For each final state $q\in F$, there is the
  production
  \begin{enuletter}
    \item $[qA]_1\rightarrow\epsilon$
  \end{enuletter}    
\end{itemize}

\noindent
The productions from a variable $[qA]_0$
are similar to those used in the standard conversion from
pda's (accepting by empty stack) to context-free 
grammars.\footnote{In that case, variables of the form $[qAp]$
are used, 
where $p$ represents one possible ``exit'' from the
segment from $[qA]$. In the case under
consideration, there is at most one possible exit,
namely $\exit[qA]$.}
The productions from modes $[qA]_1$ are used to guess
that at some point the computation will stop in a final
state. For example, for the push move 
$(p,\push(B))\in\delta(q,\epsilon,A)$,
we can guess that the acceptance will be reached in the 
segment of computation which starts from the mode
$[pB]$ (hence, ending the computation before reaching
the same stack level as in the starting mode $[qA]$, 
see production (a)), or  after that segment is
completed (production (c)).

\noindent
In order to show that the grammar $G$ is equivalent
to $M$, it is useful to prove the following lemma:

\begin{lemma}\label{lemma:prod}
  For each mode $[qA]$, $x\in a^*$, the following hold:
  \begin{enumerate}
  \item\label{prod0}
  $[qA]_0\genera x$ if and only if
  $(q,x,A)\vdashstar(\exit[qA],\epsilon,\epsilon)$.
  \item\label{prod1}
  $[qA]_1\genera x$ if and only if
  $(q,x,A)\vdashstar(q',\epsilon,\gamma)$, for
  some $q'\in F$, $\gamma\in\Gamma^+$.
  \end{enumerate}
\end{lemma}
\begin{proof}
To prove (1), we show by induction that for each 
integer $k\geq 1$, $[qA]_0\GENERA{k}x$ if and only if
$(q,x,A)\VDASH{k}(\exit[qA],\epsilon, \epsilon)$.

First of all, we observe that the case $k=1$, which
corresponds to productions (d) and to pop moves, is
trivial. For the inductive step,
we consider three subcases, depending on the move
allowed from the mode $[qA]$.

\begin{itemize}

\item $\delta(q,\epsilon,A)=\set{(p,\push(B))}$:\\
  Let $q'=\exit[pB]$ and suppose that $[qA]_0\GENERA{k}x$.
  Then, $[qA]_0\GENERA{}[pB]_0[q'A]_0$,
  $[pB]_0\GENERA{k'}x'$, $[q'A]_0\GENERA{k''}x''$,
  for some $k',k''>0$, $x',x''$ such that $k'+k''=k-1$ and
  $x'x''=x$.
  By the induction hypothesis $(p,x',B)\VDASH{k'}(q',\epsilon,\epsilon)$
  and $(q',x'',A)\VDASH{k''}(\exit[q'A],\epsilon,\epsilon)$.
  As observed above, $\exit[q'A]$ coincides with $\exit[qA]$.
  Hence:
  $(q,x,A)\vdash(p,x'x'',BA)\VDASH{k'}(q',x'',A)
   \VDASH{k''}(\exit[qA],\epsilon,\epsilon)$, that implies 
  $(q,x,A)\VDASH{k}(\exit[qA],\epsilon,\epsilon)$.
  In a similar way, the converse can be proved.

\item $\delta(q,\epsilon,A)=\set{(p,\pop)}$:
  impossible for $k>1$.

\item $\delta(q,\sigma,A)=\set{(p,\read)}$, with
      $\sigma\in\set{a,\epsilon}$:\\
  By production (e), $[qA]_0\GENERA{}\sigma[pA]_0$.
  Furthermore, $(q,\sigma,A)\vdash(p,\epsilon,A)$.
  By the induction hypothesis, for each terminal string $y$,
  $[pA]_0\GENERA{k-1}y$ if and only if
  $(p,y,A)\VDASH{k-1}(\exit[pA],\epsilon,A)$.
  The proof can be easily completed, by
  choosing $y$ such that $x=\sigma y$, and by observing
  that $\exit[pA]$ must coincide with $\exit[qA]$.
\end{itemize}

\smallskip

(2) Let us start by proving the ``only if'' part,
  by induction on the length $k$ of the derivation 
  $[qA]_1\GENERA{k}x$.

For the basis, $k=1$, the derivation must consists only of
a production of the form (f).
This implies that $q\in F$. Hence the corresponding computation
is trivial and consists only of the configuration
$(q,\epsilon,A)$.
For $k>1$ we consider different subcases, depending on
the first used production:
\begin{itemize}
\item Production (a), namely $[qA]_1\rightarrow [pB]_1$, with
  $\delta(q,\epsilon,A)=\set{(p,\push(B))}$:\\
  $[pB]_1\GENERA{k-1}x$ and, by inductive hypothesis
  $(p,x,B)\vdashstar(q',\epsilon,\gamma)$, for some
  $q'\in F$, $\gamma\in\Gamma^+$. Hence:
  $(q,x,A)\vdash(p,x,BA)\vdashstar(q',\epsilon,\gamma A)$.

\item Production (c), namely $[qA]_1\rightarrow [pB]_0[q'A]_1$, with
  $q'=\exit[pB]$ and $\delta(q,\epsilon,A)=\set{(p,\push(B))}$:\\
  $[pB]_0\GENERA{k'}x'$, $[q'A]_1\GENERA{k''}x''$, with
  $x'x''=x$, $k'+k''=k-1$. From (1) we get that
  $(p,x',B)\vdashstar(q',\epsilon,\epsilon)$ and, from the
  inductive hypothesis, $(q',x'',A) \vdashstar(q'',\epsilon,\gamma)$,
  with $q''\in F$, $\gamma\in\Gamma^+$. Hence:
  $(q,x,A)\vdash(p,x'x'',BA)\vdashstar(q',x'',A)\vdashstar
  (q'',\epsilon,\gamma)$.

\item Production (e), namely $[qA]_1\rightarrow\sigma[pA]_1$,
  with $\sigma\in\set{a,\epsilon}$, $x=\sigma y$, 
  and $\delta(q,\sigma,A)=\set{(p,\read)}$:\\
  $[pA]_1\GENERA{k-1}y$ and, by inductive hypothesis,
  $(p,y,A)\vdashstar(q',\epsilon,\gamma)$, for some
  $q'\in F$, $\gamma\in\Gamma^+$. Hence:
  $(q,x,A)\vdash(p,y,A)\vdashstar(q',\epsilon,\gamma)$.
\end{itemize}

We now prove the ``if'' part, by induction of the number $k$ of
moves in a computation $(q,x,A)\VDASH{k}(q',\epsilon,\gamma)$,
with $q'\in F$, $\gamma\in\Gamma^+$.

If $k=0$ then $q=q'$ and $x=\epsilon$. The trivial computation
is simulated by the derivation consisting only of the production
(f).

For $k>0$, we consider different subcases, depending on the
first move of the automaton:
\begin{itemize}
\item $\delta(q,\epsilon,A)=\set{(p,\push(B))}$:\\
  $(q,x,A)\vdash(p,x,BA)\VDASH{k-1}(q',\epsilon,\gamma)$.
  Because $\gamma$ is not empty, during the given computation
  the symbol $A$ cannot be removed from the stack.
  Hence $\gamma=\gamma'A$, for some $\gamma'\in\Gamma^*$,
  and $(p,x,B)\VDASH{k-1}(q',\epsilon,\gamma')$.
  
  If $\gamma'=\epsilon$ then $q'=\exit[pB]$ and, by (1),
  $[pB]_0\genera x$.
  Hence
  $[qA]_1\GENERA{}[pB]_0[q'A]_1\genera x[q'A]_1\GENERA{}x$
  (since $q'\in F$, in the last step the production (f) is used).
  
  On the other hand, if $\gamma'\neq\epsilon$, then
  by the inductive hypothesis, it turns out that
  $[pB]_1\genera x$. Hence, using production (a),
  $[qA]_1\GENERA{}[pB]_1\genera x$.

\item $\delta(q,\epsilon,A)=\set{(p,\pop)}$:\\
  This case is not possible because it
  should imply $k=1$, $x=\epsilon$, $p\in F$, and $\gamma$ empty.

\item $\delta(q,\sigma,A)=\set{(p,\read)}$, with
  $\sigma\in\set{a,\epsilon}$, $x=\sigma y$, $y\in a^*$:\\
  $(q,\sigma y,A)\vdash(p,y,A)\VDASH{k-1}(q',\epsilon,\gamma)$.
  By inductive hypothesis $[pA]\genera y$. Hence:
  $[qA]_1\GENERA{}\sigma[pA]_1\genera\sigma y=x$.    
\end{itemize}

\end{proof}

As a consequence of Lemma~\ref{lemma:prod}, it turns out
that, for each $x\in a^*$, $[q_0Z_0]_1\genera x$ 
if and only if $x$ is accepted by $M$.
Hence, we get the following result:

\begin{theorem}
  For any unary deterministic pushdown automaton $M$ in normal
  form, with $n$ states and $m$ pushdown symbols, there
  exists an equivalent context-free grammar with at most
  $2mn$ variables, such that the right hand side
  of each production contains at most two symbols.
\end{theorem}

Finally, we can observe that from the grammar $G$ above
defined, it is easy to get
a grammar in Chomsky normal formal, accepting
$L(M)-\set{\epsilon}$. This can require one more variable.

\section{Immediate acceptance/rejection}
\label{sec:imm}

Because dpda's can perform $\epsilon$-moves, 
in order to decide whether or not an input string $w$
is accepted, it is not enough to consider only the configuration
reached immediately after reading the last symbol of $w$: 
even the configurations reachable in the further steps, via
$\epsilon$-moves, must be taken into account.
In this section we show how to modify a unary pda, accepting
by final states, in order to be able to decide the acceptance 
or the rejection of an input string $w$, just considering the configuration reached
immediately after reading the last symbol of $w$.
This result will be useful for a construction presented in 
Section~\ref{sec:complex}.\footnote{We remind that as observed 
in Section~\ref{sec:prel}, 
in the unary case we can consider, without increasing the size, 
loop-free dpda's.}

More precisely, let us consider a unary (deterministic
or nondeterministic) pda 
$M=(Q,\set{a},\Gamma,\delta,q_0,Z_0,F)$ in normal
form, accepting by \emph{final states}. We define another
pda $M'$, where each transition
$(p,\read)\in\delta(q,a,A)$ of $M$ is replaced with an 
$\epsilon$-transition, postponing the reading of the symbol $a$ until a 
final state is reached or the following input symbol should
be read.

More formally, $M'=(Q',\set{a},\Gamma,\delta',q'_0,Z'_0,F')$,
with $Q'=Q\cup\tilde Q\cup\set{q'_0}$, where $\tilde Q$ is 
an isomorphic copy of $Q$ and the
transition function $\delta'$ is defined as follows,
for $q\in Q$, $\tilde q\in\tilde Q$, $\sigma\in\set{\epsilon, a}$,
$A\in\Gamma$:
\begin{itemize}
\item $\delta'(q,\epsilon,A)=\delta(q,\epsilon,A)\cup
  \set{(\tilde p,\read) \mid (p,\read)\in\delta(q,a,A)}$
\item $\delta'(q,a,A)=\emptyset$
\item $\delta'(\tilde q,\sigma,A)=
  \left\{
\begin{array}{ll}
  \set{(\tilde p, \alpha)\mid(p,\alpha)\in\delta(q,\sigma,A)}
	   &\mbox{if $q\notin F$}\\
  \set{(q,\read)}&\mbox{if $q\in F$ and $\sigma=a$}\\
  \emptyset&\mbox{otherwise}
\end{array}\right.$
\item $\delta'(q'_0,\epsilon,Z_0)=\set{(q_0,\read)}$
\end{itemize}
Intuitively, the states in $\tilde Q$ are used to remember the
debt of one read operation. The debt is paid when a final state is
reached. However, if in the original pda $M$ the read of a 
further symbol must be performed,
before reaching a final state, then in $M'$ a read is executed, 
without canceling the debt.

The new initial state $q'_0$ is useful when $q_0$ is not
accepting, but the empty word must be accepted, i.e., in the original automaton there is a sequence of
transitions leading from $q_0$ to a final state, without
consuming any input symbol. Hence:
\[
F'=  \left\{
\begin{array}{ll}
  F\cup\set{q'_0}&\mbox{if $\epsilon$ is accepted by $M$}\\
  F&\mbox{otherwise}.
\end{array}\right.
\]
Because final states (with the possible exception of $q'_0$) can
be reached only with moves that consume an input symbol,
we can conclude that $M'$ satisfies the required property of
accepting input strings immediately after reading the last symbol.
In order to prove that $M'$ is equivalent to $M$, the following
lemma is useful (the transition relations between 
configurations are marked with the names of the considered
pda's):

\begin{lemma}\label{lemma:M}
  For each $k\geq 0$, $q\in Q$, $\alpha\in\Gamma^*$:
  (a) $(q_0,a^k,Z_0)\vdashstarM(q,\epsilon,\alpha)$
  if and only if
  (b) $(q'_0,a^k,Z_0)\vdashstarP(q,\epsilon,\alpha)$
  or (c) $(q'_0,a^{k-1},Z_0)\vdashstarP(\tilde q,\epsilon,\alpha)$.
  Furthermore, if  
  $(q_0,a^k,Z_0)\vdashstarM(p,\epsilon,\beta)\vdashstarM(q,\epsilon,\alpha)$,
  for some $p\in F$, $\beta\in\Gamma^*$, then (b) holds.
\end{lemma}
\begin{proof}
The lemma can be proved by induction on the length of the derivations, and
by observing that for $q\in F$, (c) implies (b).
Because the proof is very technical and it involves only standard
arguments, it is omitted.
\end{proof}

As consequence of the previous construction and of
Lemma~\ref{lemma:M}, we get that $M$ and $M'$ are equivalent,
and hence:

\begin{theorem}\label{th:imm}
For each unary pda $M$ in normal form with $n$ states, accepting
by final states, there exists an equivalent pda $M'$ in normal
form with $2n+1$ states and the same pushdown alphabet as $M$ such
that each input string $w$ is accepted if and only if the state
reached immediately after reading the last symbol of $w$ is final.
Furthermore, if $M$ is deterministic then $M'$ is deterministic, too.
\end{theorem}

\section{Languages with complex dpda's}
\label{sec:complex}

In Section~\ref{sec:simulation}, we proved
that dpda's can be exponentially more succinct
than finite automata.
In this section we show the existence of languages for which
this dramatic reduction of the descriptional complexity cannot
be achieved.
More precisely, we prove that for each integer $m$
there exists a unary $2^m$-cyclic language $B_m$ such that
the size of each dpda accepting it is exponential in $m$.

Let us start by introducing the definition of the language
$B_m$. To this aim, we first recall that a 
\emph{de Bruijn word}~\cite{de46} of order $m$ on $\set{0,1}$
is a word $w_m$ of length $2^m+m-1$ such that each string of
length $m$ is a factor of $w_m$ occurring in $w_m$
exactly one time. Furthermore, the suffix and the prefix
of length $m-1$ of $w_m$ coincide.

We consider the following language:\footnote{The same
language was considered in~\cite{BC04} for a different
problem.}
\[
B_m=\set{a^k\mid\mbox{~the $(k \mod' 2^m)$th letter
  of $w_m$ is $1$}},
\]
where
$
x\mod'y=  \left\{
\begin{array}{ll}
  x\mod y&\mbox{if $x\mod y>0$}\\
  y&\mbox{otherwise.}
\end{array}\right.
$
\\
For example, $w_3=0001011100$ and 
$B_3=\set{a^0,a^4,a^6,a^7}\set{a^8}^*$.

By definition and by the above mentioned properties of de 
Bruijn words, $B_m$ is a properly $2^m$-cyclic unary language. 
Hence, the minimal 1dfa accepting it has exactly $2^m$ states
(actually, by Theorem 9 in~\cite{MP00}, this number of
states is required even by each 2nfa accepting $B_m$).
We show that even the size of each dpda accepting $B_m$
must be exponential in $m$. More precisely:

\begin{theorem}\label{th:complex}
  There is a constant $d$, such that for each $m>0$ the
  size of any dpda accepting $B_m$ is at least
  $d\frac{2^m}{m^2}$.
\end{theorem}
\begin{proof}
  Let us consider a dpda $M$ of size $s$ accepting $B_m$.
  We will show that from $M$ it is possible to build a 
  grammar with $O(sm)$ variables generating the language 
  which consists only of the word $w_m$. Hence, the result
  will follow from a lower bound presented in~\cite{DPS02}, 
  related to the generation of $w_m$.

  First of all, by Theorem~\ref{th:imm}, from $M$ it is possible 
  to get an equivalent
  dpda $M'$ of size $O(s)$, such that $M'$ is
  able to accept or 
  reject each string $a^k$ immediately after reading the 
  $k$th letter of the input.
  
  We also consider a 1dfa $A$ accepting the
  language $L$ which consists of all strings $x$ on the alphabet
  $\set{0,1}$, such that $x=yw$, where $w$ is the suffix
  of length $m$ of $w_m$, and $w$ is not a proper factor of
  $x$, i.e., $x=x'w$, and $x=x''ww'$ implies $w'=\epsilon$.
  Note that $A$ can be implemented 
  with $m+1$ states.
  The automaton
  $A$ will be used in the following to modify the
  control of $M'$, in order to force it to accept only
  the string $a^{2^m+m-1}$.
  
  To this aim, we describe a new dpda $M''$.
  Each state of $M''$ simulates one state of $M'$ and one
  state of $A$. The initial state of $M''$ is the pair
  of the initial states of $M'$ and $A$.
  $M''$ simulates $M'$ moves step by step. When a transition
  which reads an input symbol is simulated, then $M''$
  simulates also one move of $A$ on input $\sigma\in\set{0,1}$,
  where $\sigma = 1$ if the transition of $M'$
  leads to an accepting state, $0$ otherwise.
  In this way, the automaton $A$ will finally receive 
  as input the word $w_m$.
  When the simulation reaches the accepting state of $A$,
  namely the end of $w_m$ has been reached,
  $M''$ stops and accepts. Thus, the only string accepted
  by $M''$ is $a^{2^m+m-1}$.
  
  Using the construction  presented in Section~\ref{sec:grammar}, 
  we can build a context-free grammar $G$ equivalent
  to $M''$. We modify the productions of $G$
  that correspond to operations which consume input
  symbols: each production $[qA]_b\rightarrow a[pA]_b$
  is replaced by $[qA]_b\rightarrow 1[pA]_b$ if
  $p$ corresponds to a final state of $M'$, and by 
  $[qA]_b\rightarrow 0[pA]_b$ otherwise. It is easy
  to observe that the grammar $G'$ so obtained generates
  the language $\set{w_m}$.
  Furthermore, the size of $G'$ is bounded by $ksm$, 
  for some constant $k$.
  By a result presented in~\cite{DPS02} (based on a lower bound
  from~\cite{Al90}), the number of variables
  of $G'$ must be at least $c\frac{2^m}{m}$ for some
  constant $c$.
  Hence, from $ksm\geq c\frac{2^m}{m}$, we finally get that
  the size of the original dpda $M$ must be at least
  $d\frac{2^m}{m^2}$ for some constant $d$.
\end{proof}

\section*{Acknowledgment}
I would like to thank the anonymous referees
for their valuable comments and suggestions.

\end{document}